\documentclass[a4paper,12pt]{article}
\usepackage{graphics,graphicx}
\usepackage{amsmath}
\usepackage{latexsym}
\usepackage{setspace}
\usepackage{color}
\usepackage[margin=1in]{geometry}
\usepackage{times}
\usepackage{txfonts}
\usepackage{mathptmx}
\usepackage{subfig}
\usepackage{authblk}
\usepackage[comma,numbers,round]{natbib} 

\makeatletter
\renewcommand{\@biblabel}[1]{\quad#1.}
\makeatother

\title{\huge{A dynamic stress model explains the delayed drug effect in artemisinin treatment of \emph{Plasmodium falciparum}}}

\author[1]{Pengxing Cao}
\author[2]{Nectarios Klonis}
\author[3]{Sophie Zaloumis}
\author[2]{Con Dogovski}
\author[2]{Stanley C. Xie}
\author[4]{Sompob Saralamba}
\author[4]{Lisa J. White}
\author[3,5]{Freya J. I. Fowkes}
\author[2]{Leann Tilley}
\author[3]{Julie A. Simpson}
\author[1,3,6]{James M. McCaw\thanks{Correspondence: jamesm@unimelb.edu.au}}
\affil[1]{School of Mathematics and Statistics, The University of Melbourne, Melbourne, Australia.}
\affil[2]{Department of Biochemistry and Molecular Biology and Australian Research Council Centre of Excellence for Coherent X-Ray Science, Bio21 Molecular Science and Biotechnology Institute, University of Melbourne, Melbourne, Australia.}
\affil[3]{Centre for Epidemiology and Biostatistics, Melbourne School of Population and Global Health, The University of Melbourne, Melbourne, Australia.}
\affil[4]{Mahidol-Oxford Tropical Medicine Research Unit, Faculty of Tropical Medicine, Mahidol University, Rajthevee, Bangkok, Thailand.}
\affil[5]{Burnet Institute, Melbourne, Australia.}
\affil[6]{Modelling and Simulation, Infection and Immunity Theme, Murdoch Childrens Research Institute, The Royal Children's Hospital, Parkville, Victoria, Australia.}

\begin{document}

\date{}
\maketitle
\vspace{1.5cm}


\thispagestyle{empty}

\newpage

\section*{Abstract}

Artemisinin resistance constitutes a major threat to the continued success of control programs for malaria. With alternative antimalarial drugs not yet available, improving our understanding of how artemisinin-based drugs act and how resistance manifests is essential to enable optimisation of dosing regimens in order to prolong the lifespan of current first-line treatment options. Here, through introduction of a novel model of the dynamics of the parasites' response to drug, we explore how artemisinin-based therapies may be adjusted to maintain efficacy and how artemisinin resistance may manifest and be overcome. We introduce a dynamic mathematical model, extending on the traditional pharmacokinetic-pharmacodynamic framework, to capture the time-dependent development of a stress response in parasites. We fit the model to \emph{in vitro} data and establish that the parasites' stress response explains the recently identified complex interplay between drug concentration, exposure time and parasite viability. Our model demonstrates that the previously reported hypersensitivity of early ring stage parasites of the 3D7 strain to dihydroartemisinin (DHA) is primarily due to the rapid development of stress, rather than any change in the maximum achievable killing rate. Of direct clinical relevance, we demonstrate that the complex temporal features of artemisinin action observed \emph{in vitro} have a significant impact on predictions of \emph{in vivo} parasite clearance using PK--PD models. Given the important role that such models play in the design and evaluation of clinical trials for alternative drug dosing regimens, our model contributes an enhanced predictive platform for the continued efforts to minimise the burden of malaria.

\[\]
{\bf Keywords: artemisinin action, \emph{Plasmodium falciparum}, drug exposure time, dynamic model}

\newpage
\section*{Introduction}

\emph{Plasmodium falciparum} malaria is a major vector-borne parasitic disease affecting over 200~million people annually \cite{WHOrep2015}. Over the past two decades artemisinin-based therapies, used as the first line treatment against falciparum malaria, have been shown to be highly effective. Their widescale distribution (approximately 390~million treatment courses delivered annually) has been instrumental in achieving a dramatic reduction in morbidity and mortality through both individual-level clinical and public health benefits \cite{WHOrep2015}. Worryingly, over the past decade \emph{P. falciparum} parasites resistant to artemisinin derivatives---originally defined via a clinical phenotype of delayed clearance following treatment and now characterised by presence of the K13 mutation---have begun to emerge and spread across South East Asia \cite{Dondorpetal2009,Phyoetal2012,Arieyetal2014,Ashleyetal2014}. With no new antimalarial drugs yet available, and alternatives unlikely to be brought to market within the the next few years, advancing our understanding of the antimalarial action of the artemisinins is essential to prolong the lifespan of the current first-line treatment for malaria.

A model-based study of clinical isolates from Pailin (Western Cambodia) by Saralamba~\emph{et al.} in 2011 demonstrated that artemisinin-resistant parasites displayed a reduced sensitivity to artesunate (an artemisinin derivative with active metabolite dihydroartemisinin) during the ring-stage of infection \cite{Saralambaetal2011}. Recent \emph{in vitro} experiments have further demonstrated that \emph{P. falciparum} exhibits a distinct stage-dependent susceptibility to artemisinin, and that resistant isolates show a reduced drug susceptibility during the very early ring stage of development \cite{Klonisetal2013,Witkowskietal2013,Dogovskietal2015}. Despite this developing understanding of the subtleties of artemisinin action and drug resistance, a major gap remains in describing the full dynamics of the host-pathogen-drug system and translating findings from the well-controlled \emph{in vitro} experimental environment to the \emph{in vivo} context.

Pharmacokinetic--pharmacodynamic (PK--PD) modelling, which integrates drug kinetics (e.g. absorption and elimination) with the dynamics of both cyclic parasite growth and drug--parasite interactions, enables the quantitative assessment of drug efficacy, determination of optimal dosing schemes and the advancement of our understanding of antimalarial action and resistance \cite{Austinetal1998,Simpsonetal2000,Hoshenetal2001,Saralambaetal2011,Zaloumisetal2012,White2013,Kayetal2013,Pateletal2013,Hodeletal2014,Johnstonetal2014,Hodeletal2016}. Over nearly 20~years of development, PK--PD models have increased significantly in complexity. Building from early models, which treated infected red blood cells as a single compartment \cite{Austinetal1998,Simpsonetal2000,Hoshenetal2001}, models have expanded to capture the different stages of the parasite life cycle in the red blood cell (ring, trophozoite, schizont) allowing the incorporation of stage-dependent drug effects \cite{Saralambaetal2011,Zaloumisetal2012,Hodeletal2016}. A feature common to all PK--PD models of artemisinin-based therapy developed to date has been the \emph{implicit assumption} that the relationship between drug concentration and the rate of parasite killing is independent of the history of exposure. The transient killing rate $k$ (i.e.\ the fraction of parasites killed by drug per unit time) has been empirically modelled by a Hill function of plasma drug concentration ($C$),
\begin{equation} 
\label{eq:I1}
k(C) = \frac{k_{max}C^{\gamma}}{{K_c}^{\gamma}+C^{\gamma}},
\end{equation}
where parameters $k_{max}$, $\gamma$ and $K_c$ are (possibly stage-dependent) fixed quantities (i.e.\ constants) \cite{Simpsonetal2014}. The killing rate varies with drug concentration in a sigmoidal manner and saturates at the maximum killing rate $k_{max}$ for high drug concentration. Under this formulation a higher drug concentration will \emph{immediately} exert a stronger killing effect. 

However, the recent \emph{in vitro} experiments of Klonis~\emph{et al.}\ \cite{Klonisetal2013} and Dogovski~\emph{et al.}\ \cite{Dogovskietal2015} have provided clear evidence that a higher drug concentration may not result in an increased killing effect and indeed may be less effective if the exposure time is shortened. They demonstrated that the fraction of parasites that remains viable (i.e.\ able to asexually reproduce and so initiate a subsequent round of blood-stage infection) does not depend solely on the applied drug concentration. Rather, viability was established to be a complex function of the \emph{cumulative} drug exposure and the initial drug concentration, manifesting as (stage-dependent) variations in the exposure time required to render parasites non-viable \cite{Klonisetal2013}. The minimum exposure time required for loss of viability was particularly extended for mid-ring stage parasites (artemisinin sensitive 3D7 laboratory strain). Antimalarial resistance corresponded to a distinct change in the susceptibility of early ring-stage parasites \cite{Dogovskietal2015}. To date, these novel properties have not been incorporated into a mechanistic model of parasite killing, indicating a requirement to extend the PK--PD modelling framework to reflect our emerging understanding of drug activity and evaluate the influence of these novel biological phenomena on the prediction of parasite clearance \cite{Simpsonetal2014}.

\begin{figure}[ht!]
\centering
\includegraphics[scale=1]{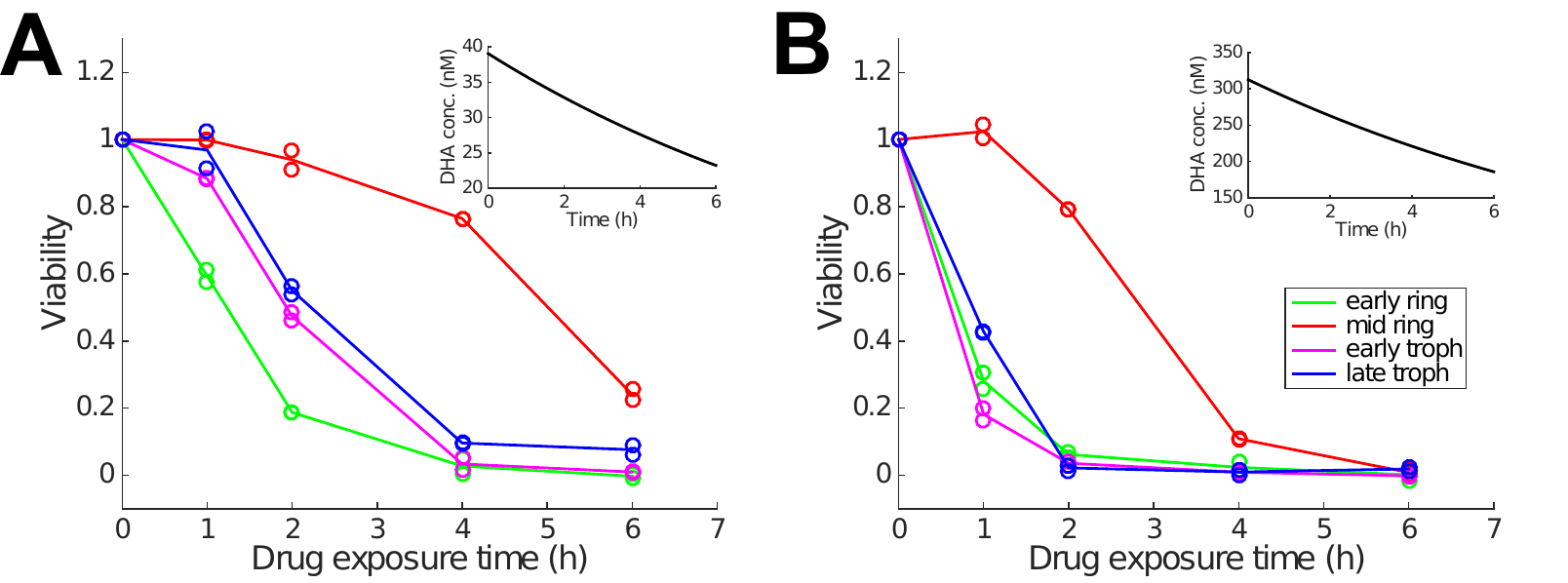}
\caption{Representative experimental data showing how the fraction of viable parasite (i.e.\ viability) changes with the duration of drug exposure for two different initial DHA concentrations (39~nM (left) and 300~nM (right)) and four different parasite life stages. Insets indicate the \emph{in vitro} decay of DHA concentration. Empty circles display the raw viability data and the curves pass through the arithmetic means of the paired data points. (Data sourced from \cite{Klonisetal2013}.)}
\label{fig:klonis_data}
\end{figure}

In this paper, we generalise the traditional model of killing (Eq.\ \ref{eq:I1}) by allowing the maximum killing rate $k_{max}$ and the half-maximal killing concentration $K_c$ to be time-dependent quantities and then fit the generalised model to viability data for the 3D7 laboratory strain available in \cite{Klonisetal2013}. By doing so, we aim to 1) show if the model is able to capture the full set of \emph{in vitro} viability data; and 2) elucidate how the artemisinin-mediated killing effect develops following drug exposure and how that development differs between the parasite life stages. These results will further imply the relative contributions of drug concentration and exposure time to the effective killing rate. Finally, we incorporate the time-dependent drug effect into a PK--PD modelling framework to evaluate its effect on \emph{in vivo} parasite killing. Complex temporal effects are anticipated to be present as the short half life of the artemisinins \emph{in vivo} is comparable to the exposure time required for effective parasite killing.

\section*{Materials and Methods}

\subsection*{\emph{In vitro} experiments and data}

In order to identify the key features that motivate the development of our model, we first review the \emph{in vitro} experimental procedure (see \cite{Klonisetal2013} for details). Cultures containing equal quantities of tightly age-synchronised \emph{P. falciparum} parasites (3D7 laboratory strain; over 80\% of parasites synchronised within a one-hour age window) were treated with a specified dose of dihydroartemisinin (DHA) for a duration of 1, 2, 4 or 6~hours before washing (to remove all drug). To quantify the effect of drug a viability assay was performed. \emph{Viable} parasites were defined as those able to reproduce and enter the next cycle of replication (thus excluding dead and dormant populations), assessed by measuring the parasitaemia ($P$) in the trophozoite stage in the \emph{following} life cycle, 48~hours later. In order to calculate the viability, parasitaemia was also measured for two special cases: the control case ($P_{control}$) where no drug was applied; and the background case ($P_{background}$) with supermaximal DHA concentration ($> 10 \times$ the 50\% lethal dose of 3 days, nM) applied for over 48 hours. Viability ($V$), a ratio and so unitless, was then given by subtracting the unviable population,
\begin{equation} 
V = \frac{P - P_{background}}{P_{control} - P_{background}}.
\end{equation}
To study stage-specific drug effects, Klonis~\emph{et al.}\ \cite{Klonisetal2013} tested four different parasite ages (by using different age-synchronised groups): 2 hours post-infection (h p.i.; early ring stage), 7.5 h p.i. (mid-ring stage), 24 h p.i. (early trophozoite stage) and 34 h p.i. (late trophozoite stage). Two examples of viability data are given in Fig.\ref{fig:klonis_data} which show the viability for different durations of drug exposure (1h, 2h, 4h, and 6h) with an initial DHA concentration of approximately 39~nM or 300~nM. Note that DHA concentration also decays \emph{in vitro} with a half life of approximately 8~hours. Experiments were performed in technical replicates for each combination of initial DHA concentration and drug exposure duration.

\subsection*{The model}

We take as our fundamental conceptualisation of antimalarial action that the drug kills or otherwise prevents parasites within infected red blood cells (iRBC) from being able to produce viable merozoites (which would go on to invade and infect other RBC at the end of the first life cycle.) We model the number of iRBC of age ~$a$ (i.e.\ RBCs that have been infected with parasites for $a$~hours) surviving drug exposure ($N(a,t)$) by the first-order partial differential equation,
\begin{equation} 
\label{eq:M1a}
\frac{\partial N(a,t)}{\partial t} + \frac{\partial N(a,t)}{\partial a} = -kN(a,t),
\end{equation}
where $k$ is a drug-induced parasite killing rate and may depend on other factors such as drug concentration, parasite life stage or even drug exposure duration (which will be explicitly indicated once we formally introduce those dependencies later). We have the boundary condition $N(0,t) = rN(48,t)$, where $r$ is the parasite multiplication factor, indicating the average number of newly-infected RBC generated from merozoites released from a single iRBC at the end of the preceding life cycle. 

The \emph{in vitro} experiments use tightly age-synchronised parasites, allowing for further simplification to an ordinary differential equation system which is sufficient for determining the time-dependency in the maximum killing rate $k_{max}$ and the half-maximal killing concentration $K_c$. We track only the number of newly infected RBC generated from parasites first exposed to drug at age $\bar{a}$ (denoted by $\bar{N}(t)$):
\begin{equation} 
\label{eq:M1b}
\frac{d\bar{N}(t)}{dt} = -k\bar{N}(t).
\end{equation} 
As mentioned in the \emph{Introduction}, the parasite killing rate $k$ is empirically modelled by a Hill function of drug concentration ($C(t)$),
\begin{equation} 
\label{eq:M2}
k(C(t)) = \frac{k_{max}C(t)^{\gamma}}{{K_c}^{\gamma}+C(t)^{\gamma}},
\end{equation}
where $k_{max}$ is the maximum killing rate, $K_c$ indicates the drug concentration at which half maximal killing, $k_{max}/2$, is achieved and $\gamma$ is the Hill coefficient. To capture the time-dependent features of the \emph{in vitro} data, we generalise the model by allowing $k_{max}$ and $K_c$ to be dependent on the duration of drug exposure. We consider the time-variation to be a function of an auxiliary modulatory variable $S(t)$ which we will refer to throughout as a general cell ``stress''. During drug exposure, parasites develop a stress response, the extent of which determines the killing effect (and thus the concentration-killing rate function, Eq.\ \ref{eq:M2}). The stress, $S(t)$, is normalised to vary between 0 and 1 (inclusive). We consider $S$ to increase in the presence of drug above some (very small) threshold level $C^*$ but decrease once drug concentration $C$ is below $C^*$. For the increase phase, we apply a simple first-order differential equation:
\begin{equation}
\label{eq:M3}
\frac{dS}{dt} = \lambda (1-S),
\end{equation}
where $\lambda$ is a rate constant which sets the time-scale for stress development. In the absence of additional experimental data, $S(t)$ is assumed to immediately reset to zero once drug concentration falls back below $C^*$. While this is sufficient to capture all available \emph{in vitro} data, we anticipate that further experimental research will allow us to more closely tie empirical determinations of the mechanisms of stress and its accumulation to our modulatory variable $S$ (with consequential changes to Eq.\ \ref{eq:M3}.)

With $S$ defined, we then model $k_{max}$ and $K_c$ as functions of $S$. $k_{max}$ is evidently positively correlated with $S$ and $K_c$ negatively correlated, indicating that as stress accumulates, the ability of the drug (at a given concentration) to kill parasites increases. In the absence of detailed experimental data, we assume these relationships are linear:
\begin{equation}
k_{max} = \alpha S,
\end{equation}
and
\begin{equation}
K_{c} = \beta_1 (1-S) + \beta_2,
\end{equation}
where $\alpha$, $\beta_1$ and $\beta_2$ are parameters to be determined.

Under this simple formulation of the model for stress accumulation and the linear relationship between stress and killing, we can solve Eq.\ \ref{eq:M3} to obtain:
\begin{equation}
\label{eq:M3a}
S(t) = 1-e^{-\lambda t},
\end{equation}
and thus
\begin{equation}
k_{max} = \alpha (1-e^{-\lambda t}),
\end{equation}
and
\begin{equation}
K_{c} = \beta_1 e^{-\lambda t} + \beta_2.
\end{equation}

Of note, when $\lambda \to \infty$ (i.e.\ the modulatory variable reaches its steady state instantaneously), $k_{max} = \alpha$ and $K_c = \beta_2$ such that our model reduces to a traditional PK--PD model with a fixed relationship between drug concentration and killing rate. 

For finite $\lambda$, $k_{max}$ and $K_c$ become functions of $S$ and so duration of exposure. In particular, for low $\lambda$, our model displays a slow development of the stress response and is thus capable of capturing a delayed reduction in viability, indicating its potential suitability for the \emph{in vitro} data shown in Fig.~\ref{fig:klonis_fit} and detailed in \cite{Klonisetal2013,Dogovskietal2015}. For simplicity, in this paper we now refer to the traditional killing rate model with constant (although perhaps stage-dependent) $k_{max}$ and $K_c$ as the \emph{stationary model} and refer to our generalised model as a \emph{dynamic stress model}. 

The dynamic stress model contains five parameters ($\lambda$, $\gamma$, $\alpha$, $\beta_1$ and $\beta_2$) to be determined by fitting to available data. These five parameters are assumed to be stage-specific, and thus the model is fitted separately to the viability data for each parasite stage. Note that, by incorporating the phenomenological model of stress through the modulatory variable $S$, we aim to develop a better understanding of how the killing rate evolves in the presence of drug. While we are as yet unable to explore the underlying mechanisms governing the development of both stress and drug action (e.g.\ the changes at the cellular or even molecular levels), we discuss possible biological interpretations further in the \emph{Discussion}.

\subsection*{Derivation of the expression for viability}

For modelling the \emph{in vitro} experiments of tightly age-synchronised parasites, we use Eq.\ \ref{eq:M1b}. The solution to Eq.\ \ref{eq:M1b} subject to an initial condition $\bar{N}(0) = \bar{N_0}$ is:
\begin{equation} 
\label{eq:M4}
\bar{N}(t) = \bar{N_0}e^{-\int_{0}^{t} k(C(\tau),S(\tau))d\tau},
\end{equation}
where we have explicitly presented the killing rate $k$ as a function of DHA concentration $C$ and stress $S$, both of which are functions of time (following initiation of drug exposure). DHA concentration $C(t)$ as a function of time due to \emph{in vitro} decay is given by
\begin{equation} 
\label{eq:M9}
C(t) = C_0 e^{-\frac{ln(2)}{t_{1/2}}t},
\end{equation} 
where $C_0$ is the initial dug concentration and the \emph{in vitro} half-life of DHA ($t_{1/2}$) was measured to be about 8 hours \cite{Yangetal2016}.

For a drug pulse with a duration of $T_d$~hours in a given stage of the parasite life cycle, the total number of iRBCs , $N_d$, at the time of data collection (during the trophozoite stage in the next life cycle) is given by 
\begin{equation} 
\label{eq:M5}
N_d = r\bar{N_0}e^{-\int_{0}^{T_d} kd\tau} + \bar{N_0}(1-e^{-\int_{0}^{T_d} kd\tau}),
\end{equation}
where $k$'s dependence on $C(\tau)$ and $S(\tau)$ is now implicit. The first term represents the number of iRBC with live parasites (having expanded by factor $r$, the parasite multiplication factor), while the second term represents the number of non-viable parasites. For the control case (no drug), the number of parasites, $N_c$, is given by Eq.\ \ref{eq:M5} with $k=0$ ($N_c = r\bar{N_0}$). For the background case (all parasites killed due to super-maximal exposure), the number of parasites, $N_b$, is given by Eq.\ \ref{eq:M5} with $k\to\infty$ ($N_b = \bar{N_0}$). Substituting these two expressions back into Eq.\ \ref{eq:M5}, we have
\begin{equation} 
\label{eq:M6}
N_d = N_ce^{-\int_{0}^{T_d} kd\tau} + N_b(1-e^{-\int_{0}^{T_d} kd\tau}),
\end{equation}
which can be rearranged to give
\begin{equation} 
\label{eq:M7}
e^{-\int_{0}^{T_d} kd\tau} = \frac{N_d-N_b}{N_c-N_b}.
\end{equation} 
The right-hand side is precisely the parasite viability ($V$) as defined in the \emph{in vitro} experiments \cite{Klonisetal2013,Dogovskietal2015}. Therefore we have
\begin{equation} 
\label{eq:M8}
V(C_0, T_d) = e^{-\int_{0}^{T_d} k(C(\tau),S(\tau))d\tau},
\end{equation} 
where the dependence of $V$ on the initial DHA concentration $C_0$ is established through Eq.\ \ref{eq:M9}. Furthermore, we identify $\phi = {\int_{0}^{T_d} k(C(\tau),S(\tau))d\tau}$ as the cumulative drug effect. We have identified the parasite viability as a function of the cumulative drug effect arising from a dynamic mechanistic model, allowing us to fit our model to available viability data, and then use the estimated parameters to perform detailed PK--PD simulations in an \emph{in vivo} context.

\subsection*{Statistical methods}

Estimates of the model parameters ($\lambda$, $\gamma$, $\alpha$, $\beta_1$, and $\beta_2$), for each parasite stage (early ring, mid-ring, early trophozoite and late trophozoite), were obtained using nonlinear mixed effect (NLME) modelling to fit Eq.\ \ref{eq:M8} separately to the viability data for each stage. For each stage, the data with different drug concentrations was fitted simultaneously. (Note that the stage-dependent estimate of $\gamma$ suggested a very limited variation (see Table~S1) and was thus fixed to be the mean of the four estimates in Table~S1 later to reduce uncertainty; main results in the paper were based on a fixed $\gamma$.) To account for the dependency between duplicate measurements, the residual error term was partitioned into between- and within-duplicate components, that were assumed to be uncorrelated and normally distributed with means of zero and variances $\sigma_b^2$ and $\sigma_w^2$, respectively. The M3 method was used to account for viability data below the quantification limit of 0.005 \cite{Ahnetal2008}. 

Model-based 95\% confidence intervals were calculated using asymptotic standard errors (square root of the inverse Fisher information) as follows: estimate $\pm$ 1.96 $\times$ asymptotic standard error. 95\% parametric bootstrap confidence intervals for the model parameters (Tables 1 and S1) and predictions (Figs. 2 and S1--S3) were calculated by: (1) generating 500 parametric bootstrap datasets by simulating from the fitted NLME model; (2) obtaining bootstrap estimates of the model parameters and predictions by re-fitting the NLME model to each parametric bootstrap dataset; and (3) calculating basic bootstrap confidence intervals for each parameter and prediction: (2 $\times$ estimate -- 97.5th percentile of bootstrap estimates,  2 $\times$ estimate -- 2.5th percentile of bootstrap estimates) \cite{Davisonetal1997}. Parametric prediction intervals for a new viability measurement (Figs. 2 and S1--S3) were calculated at the observed pulse durations and DHA concentrations by: (1) simulating 500 viability datasets from the fitted NLME model; and (2) calculating the 2.5th and 97.5th quantiles of the viability measurements simulated at each observed pulse duration and DHA concentration.

NONMEM 7.3.0 (ICON Development Solutions, Ellicott City, MD) and Perl-speaks-NONMEM $3.7.6$ \cite{Lindbometal2005} were used to perform the NLME modelling of the viability data and obtain asymptotic standard errors; and to perform the simulation-estimation procedure required to construct the 95\% parametric bootstrap confidence intervals, and the simulations necessary to calculate 95\% parametric prediction intervals. MATLAB (version 2014b; The MathWorks, Natick, MA) was used to summarise and visualise the fitting results.

\section*{Results}

\subsection*{Fitting the model to viability data}

\begin{figure}[ht!]
\centering
\includegraphics[scale=1]{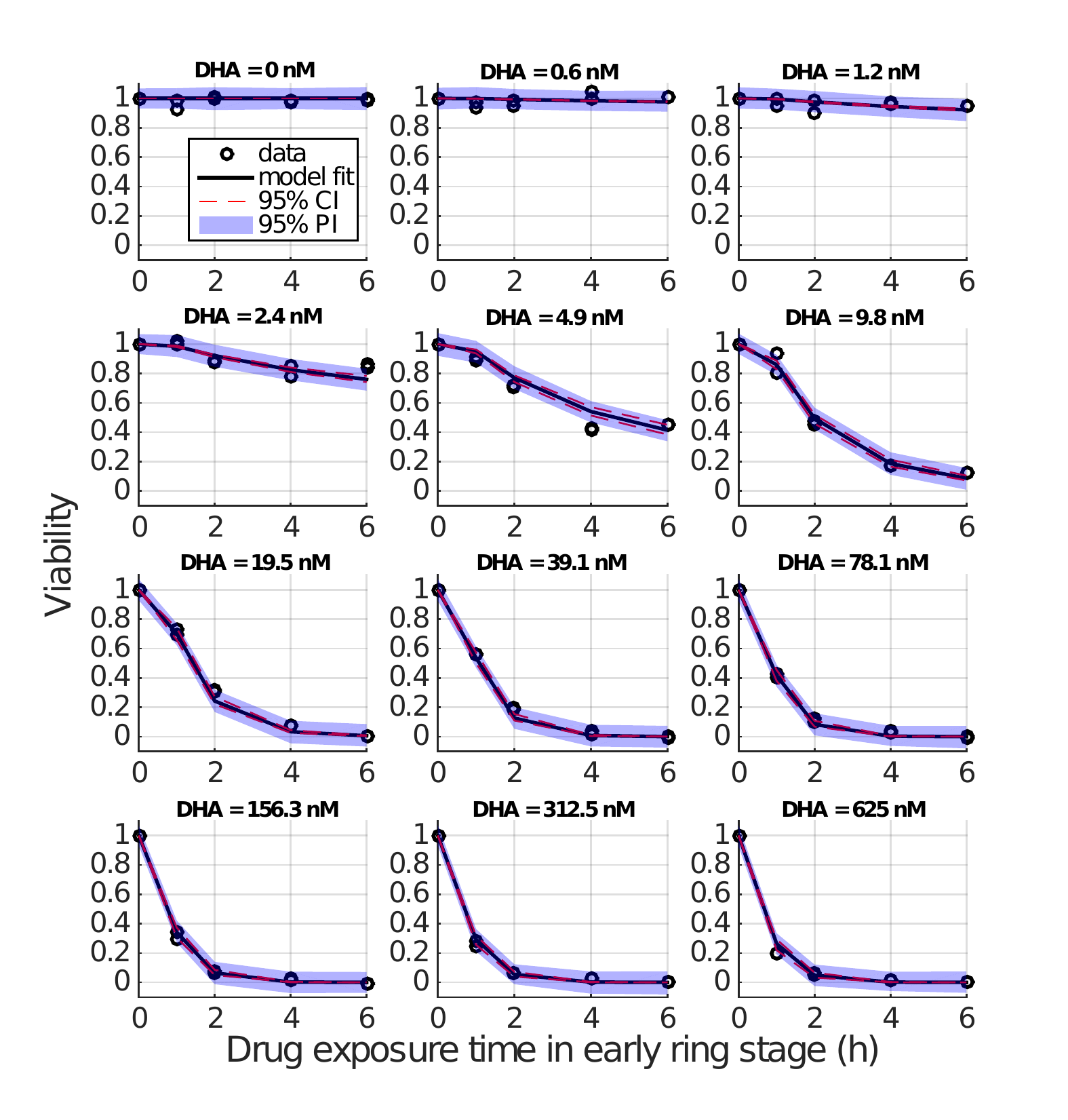}
\caption{Results of fitting the model to viability data (early ring stage). The initially applied DHA concentration is indicated in the title of each panel. Open circles (appearing in duplicate) are the repeated measures of viability by (initial) drug concentration and exposure duration. Black curves show the best-fits of the model with 95\% confidence intervals (CI) and 95\% prediction intervals (PI).}
\label{fig:klonis_fit}
\end{figure}

Fig.\ \ref{fig:klonis_fit} shows the fitting result for the early ring stage. Results for the other stages are provided in Figs.\ S1--S3. Parameter estimates and confidence intervals (CI) are given in Table~1. The model captures the data very well, in particular the dependence of viability on drug exposure time, which is the key advance we require. 

\subsection*{Drug concentration-killing rate curves and stage-dependency}

\begin{figure}[ht!]
\centering
\includegraphics[scale=1]{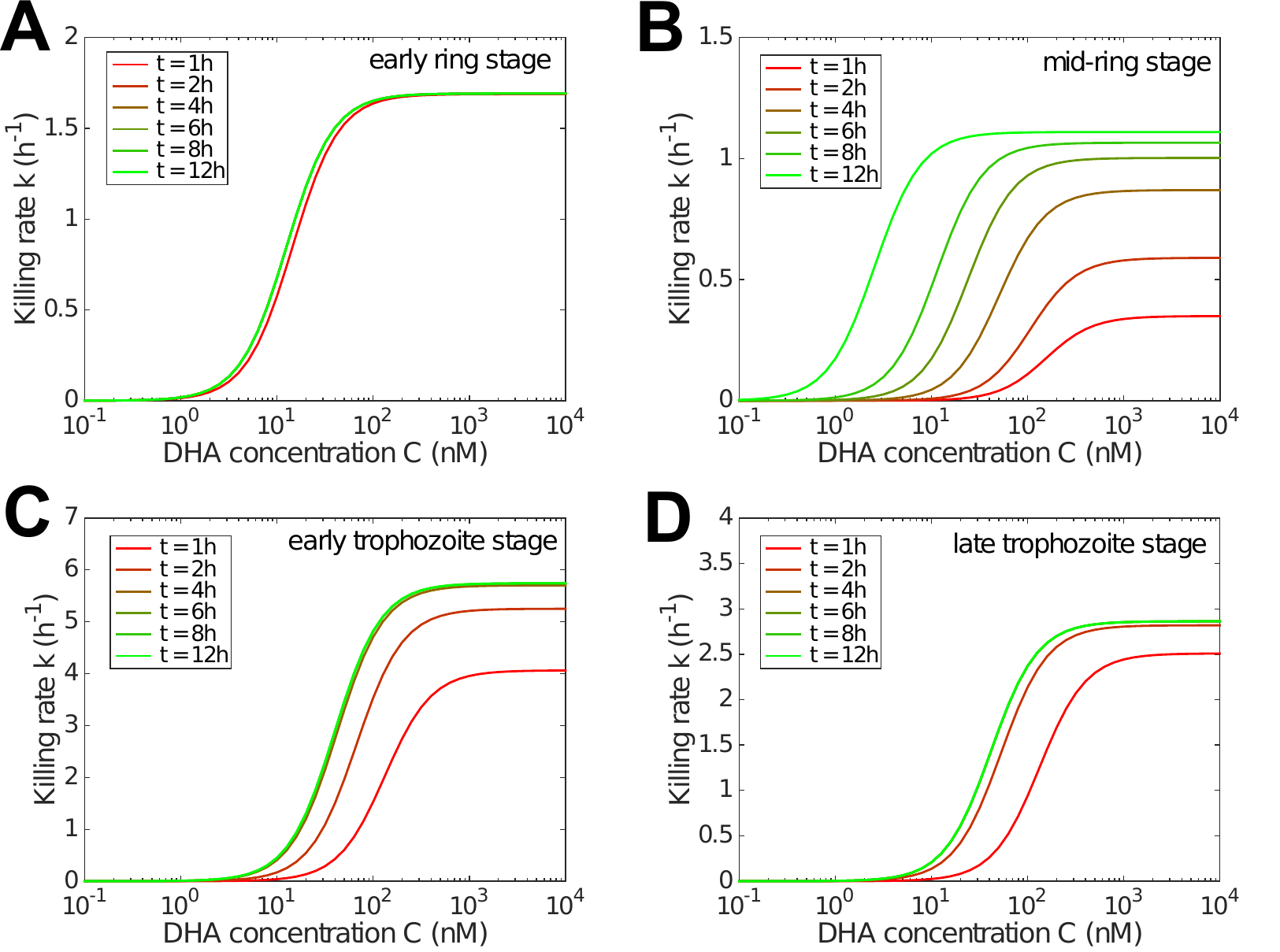}
\caption{Model results showing the evolution of the drug concentration-killing rate curve with drug exposure duration for different stages. The time after drug exposure $t$ is indicated in the legend. Note that the y-axis scale differs for different stages.}
\label{fig:killing_curves}
\end{figure}

The overall impact of parasite killing is primarily determined by the drug concentration-killing rate curve, which we now consider to be a function of exposure time, generalising the usual modelling assumption that the killing rate is an instantaneous function of drug concentration. Fig.\ \ref{fig:killing_curves} shows the modelled evolution of the concentration-killing rate curve for the four different life-stage of the parasites used in Fig.\ \ref{fig:klonis_data}. Except for the early ring stage, for which the curve reaches its steady state very quickly, the delayed process of approaching the steady-state killing rate curve for the other three stages is biologically significant, in particular for the mid-ring stage where a very strong delay is observed.  

\begin{figure}[ht!]
\centering
\includegraphics[scale=1]{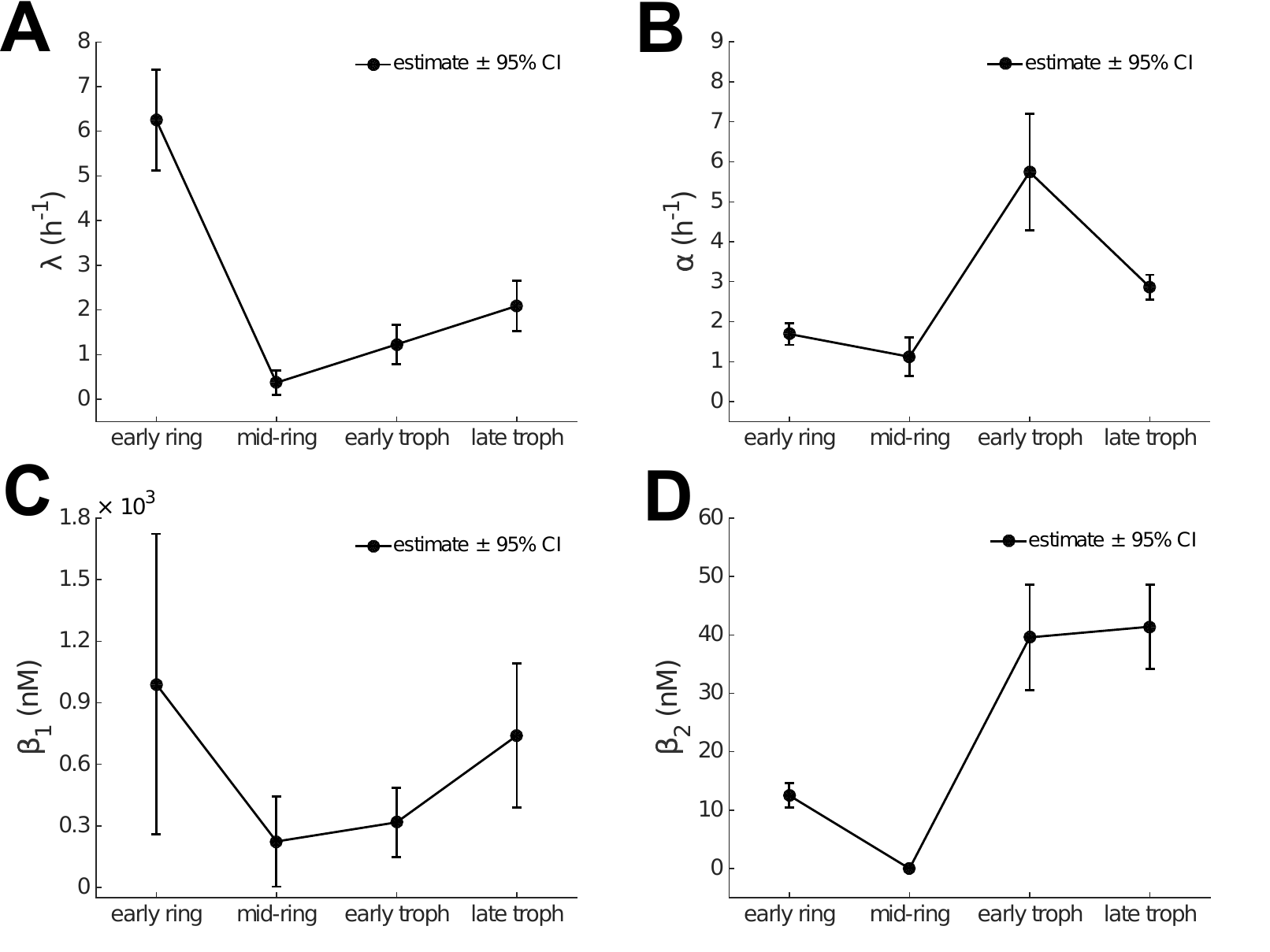}
\caption{Dependence of model parameters on parasite life stage. The parameter estimates are provided in Table~1. Error-bars show the model-based 95\% CI.}
\label{fig:params}
\end{figure}

Fig.\ \ref{fig:params} shows the estimates and 95\% confidence intervals for the four model parameters $\lambda$, $\alpha$, $\beta_1$ and $\beta_2$ by stage. The delay in the evolution of the drug concentration-killing rate curve is primarily determined by the parameter $\lambda$ (Fig.\ \ref{fig:params}A, Table~1) which reflects the accumulation rate for stress (Eq.\ \ref{eq:M3}). For the early ring stage, $\lambda = 6.25 \rm\ h^{-1}$ and thus $S(1~h) > 0.99$, indicating that early rings rapidly succumb to drug exposure. In contrast, the rate of accumulation of stress for mid-rings is much lower ($\lambda = 0.37 \rm\ h^{-1}$) and it would take over 12 hours of continued exposure to drug for $S$ to exceed 0.99. Early and late trophozoite stages display similar characteristics in terms of the rate of accumulation of stress (Fig.\ \ref{fig:killing_curves}). The maximum killing rate $\alpha$ (Fig.\ \ref{fig:params}B) shows a significant reduction for (early and mid) rings compared to (early and late) trophozoites, suggesting that young parasites may be more resistant to DHA than mature parasites once (or even when) the killing effect has reached a steady state. The parameters related to the concentration required to achieve the half-maximal killing rate, $\beta_1$ and $\beta_2$, also exhibit stage-specificity (Fig.\ \ref{fig:params}C,D). In particular, the stationary half-maximal killing concentration ($\beta_2$, Fig.\ \ref{fig:params}D) shows that ring-stage parasites exhibit a higher sensitivity to drug at steady state than trophozoites. However, it must be remembered that, particularly for mid-ring stages, the progress towards that steady-state (governed by $\lambda$) is slow, and the \emph{net effect} of the dynamics of drug-induced killing is best understood through Fig.\ \ref{fig:killing_curves}.


\subsection*{Incorporating the delayed drug effect into PK-PD modelling}

Having established the applicability of our model to the \emph{in vitro} data, we now consider the potential implications for the \emph{in vivo} application of artemisinin-based medication. We do so by incorporating the time-dependency on the killing rate into the general PK--PD framework (Eq.\ \ref{eq:M1a}) where we allow for realistic drug pharmacokinetics (PK) and a general age-structure for the parasites.

\begin{figure}[ht!]
\centering
\includegraphics[scale=1]{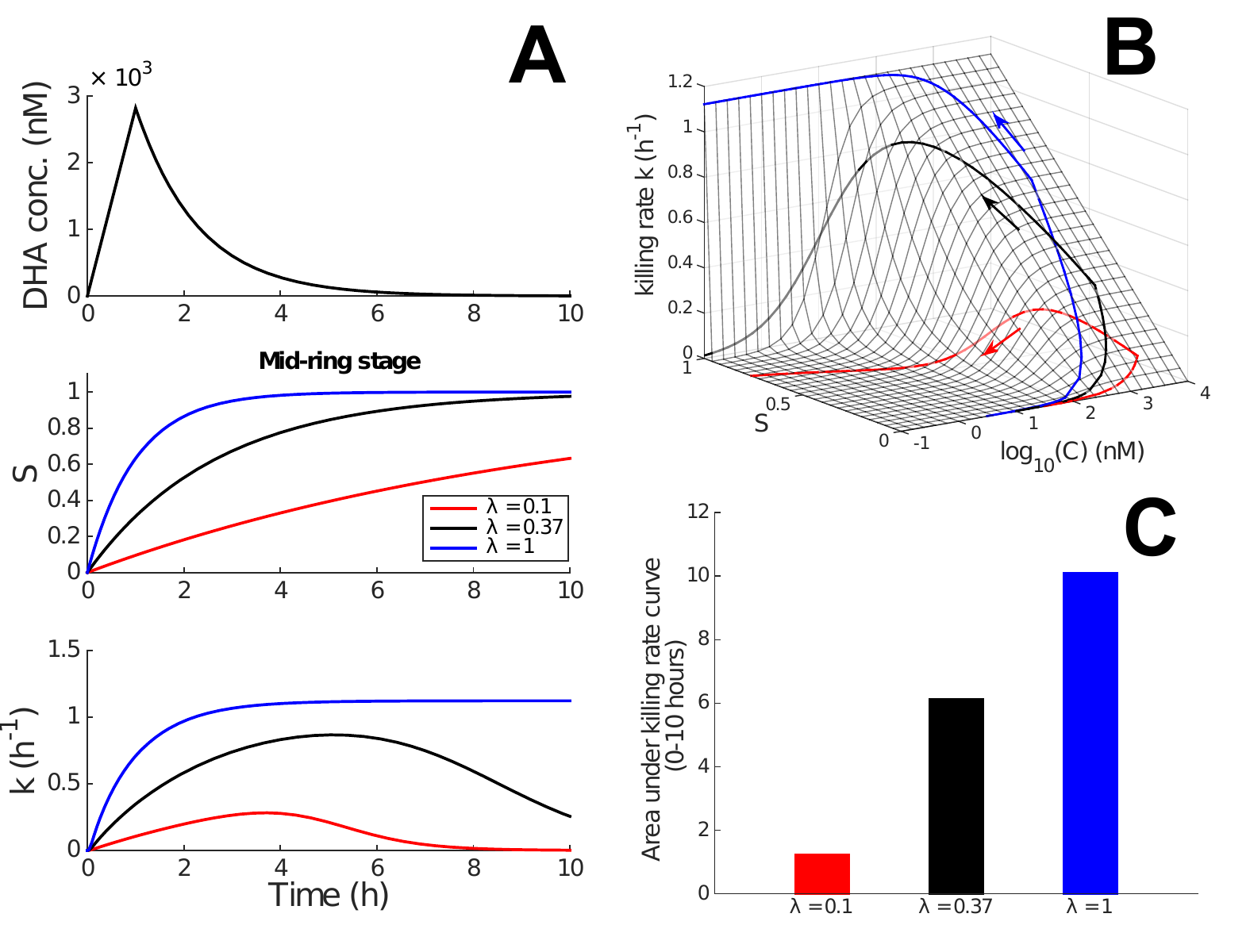}
\caption{Incorporation of the time-dependent killing rate into the PK--PD model. We study the mid-ring stage for illustrative purposes. Parameter values are taken from Table~1. Panel {\bf A} shows the simulated \emph{in vivo} DHA concentration profile (upper panel), the kinetics for the modulatory stress variable $S$ (middle panel, black curve, $\lambda=0.37$) and the transient killing rate $k$ (lower panel, black curve, $\lambda=0.37$) induced by the drug pulse. The middle and lower panels also show how $S$ and $k$ evolve if $\lambda$ is higher (blue) or lower (red). Panel {\bf B} presents the killing rate surface as a function of DHA concentration, $C$, and the stress, $S$, and the projection of the trajectory of the effective killing rate (i.e.\ a projection of the curves in Panel {\bf A} (lower panel)) onto the surface. Panel {\bf C} shows the area under the killing rate curve, an indication of the total amount of killing achievable over the course of the drug pulse.}
\label{fig:PKPD-er}
\end{figure}

We begin by considering mid-ring stage parasites treated with a single dose of artesunate (2mg/kg). The plasma DHA concentration $C(t)$ displays biphasic behaviour \cite{Saralambaetal2011}:
\begin{equation}
\frac{dC}{dt} = \left\{\def\arraystretch{1.2}%
  \begin{array}{@{}c@{\quad}l@{}}
    \frac{C_{max}}{t_m}, & 0 \leq t < t_m \\
    -\frac{ln(2)}{t_{1/2}} C, & t \geq t_m\\
  \end{array}\right.
\label{eq:PKinvivo}
\end{equation}
where $C_{max}$ is the maximum achievable concentration and $t_m$ indicates the time at which that maximum concentration is achieved. Note that the half-life, $t_{1/2}$, refers to the \emph{in vivo} half-life of drug, which is much smaller than that measured \emph{in vitro} \cite{Yangetal2016}. $C_{max} = 2820\ \rm nM$, $t_m = 1\ \rm h$ and $t_{1/2} = 0.9\ \rm h$ as per \cite{Dondorpetal2009,Yangetal2016}. The simulated PK data is shown in Fig.\ \ref{fig:PKPD-er}A (upper panel). 

The middle and lower panels of Fig.\ 5A shows the time series for the stress, $S$, and killing rate, $k$, as a result of the changing DHA concentration. The black curves are generated using $\lambda = 0.37\ \rm h^{-1}$, the best-fit estimate from fitting the model to the \emph{in vitro} data for the mid-ring stage (Table~1). Decreasing $\lambda$ will delay the increase of $S$ and in turn lead to a smaller and shortened killing rate profile (red curves in Fig.\ 5A), while increasing $\lambda$ will do the opposite (blue curves in Fig.\ 5A). We thus consider a lowering of $\lambda$ as a potential manifestation of (stage-specific) artemisinin-resistance. Conversely, $\lambda = 1$ implies that parasites accumulate stress rapidly and are rendered non-viable at the stationary rate (given $C(t)$) soon after initiation of the drug pulse.

Under the model, the killing rate $k$ is now a function of two variables (as clearly shown by Eq.\ \ref{eq:M8}), the drug concentration $C(t)$ and the modulatory stress $S(t)$. We can represent this graphically by displaying the killing rate as a trajectory on a surface in $(C,S)$-space (Fig.\ \ref{fig:PKPD-er}B). In this representation, we can clearly see the  effect of $\lambda$ on the evolution of the killing rate---the trajectory corresponding to a smaller $\lambda$ has less time (controlled by $S$ in the model) to climb up the killing rate surface even when the achievable DHA concentration remains the same.

We can also consider the net cumulative effect of the drug-pulse. As defined in \emph{Materials and Methods}, $\phi = {\int_{0}^{T_d} k(C(\tau),S(\tau))d\tau}$ represents the cumulative killing rate or cumulative drug effect, which is commonly used to indicate drug efficacy. In the \emph{in vivo} simulation, $\phi$ is simply the area under the effective killing rate curve (i.e. the area under the curve in the lower panel of Fig.\ \ref{fig:PKPD-er}A). For mid-ring parasites with $\lambda=0.37$, the cumulative drug effect (Fig.\ \ref{fig:PKPD-er}C, black bar) corresponds to a reduction in viability of approximately 99.75\% over the drug-pulse. Further numerical exploration indicates that a roughly threefold increase or decrease in $\lambda$ leads to a significant difference in the cumulative drug effect (Fig.\ \ref{fig:PKPD-er}C, blue, red bars) and in turn a few orders of magnitude difference in viability.

In summary, the results presented in Fig.\ \ref{fig:PKPD-er} indicate that the temporal drug effect significantly affects the \emph{in vivo} parasite killing and thus should be considered in model-based prediction of clinical treatment. Furthermore, the visualisation of the killing rate trajectory on the $(C,S)$-plane surface suggests a clear evolutionary strategy for the parasite to escape drug pressure, particularly given the short elimination half-life of artemisinin and its derivatives. An ability to ``outlast'' the short drug pulse provides an effective means of escape, quite distinct from any changes in susceptibility as are typically considered by a change in the maximal killing rate or drug concentration required to achieve half-maximal killing.

\begin{figure}[ht!]
\centering
\includegraphics[scale=1]{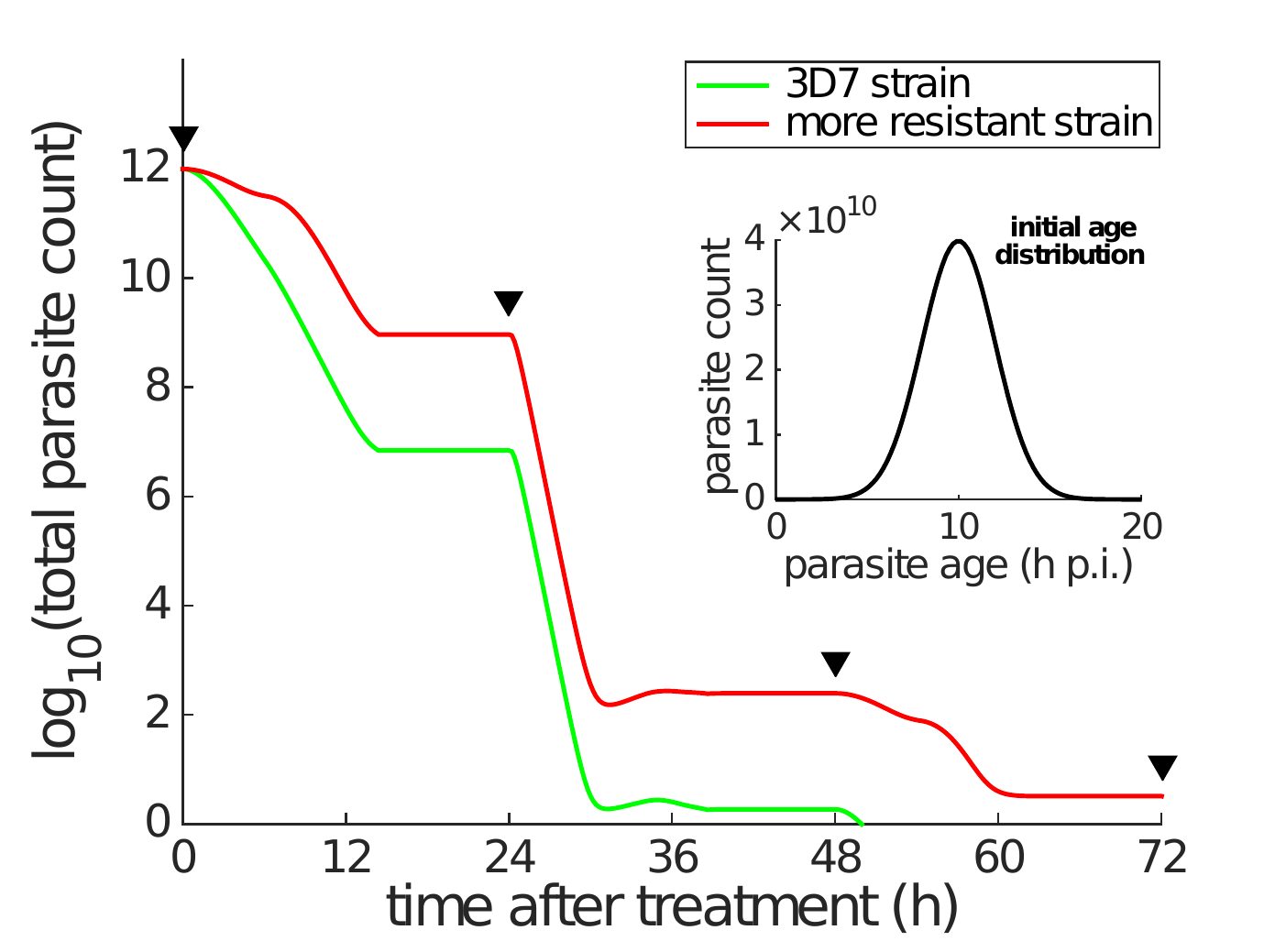}
\caption{Simulation of parasite killing under a standard treatment of 2mg/kg artesunate every 24 hours. The inset shows the age distribution of total $10^{12}$ parasites (per patient) at the start of treatment ($\sim N(10,2^2)$). The parasite multiplication factor is assumed to be 10, which means that 10 new parasites are produced once a parasite reaches 48 h p.i.. PK profile is a series of repeated DHA concentration profile every 24 hours (i.e. repeated simulations of DHA concentration profile in Fig.\ 5A upper panel). The black triangles indicate when the doses are given. The green curve corresponding to the laboratory 3D7 strain is generated using the parameters in Table~1, while the red curve is generated using the same set of parameters except for reducing $\lambda$ for mid-ring stage to be $0.1 \rm h^{-1}$ to simulate a more resistant strain. With limited information, we simply divide 48 hours' life cycle into early ring (0--6 h p.i.), mid-ring (6--26 h p.i.), early trophozoite (26--34 h p.i.) and late trophozoite (34--48 h p.i.) in the simulation. Modulatory variable $S$ is assumed to follow Eq.\ \ref{eq:M3} only when DHA concentration $C \geq 0.1 \rm\ nM$, and $S$ is immediately reset to zero when DHA concentration drops below 0.1 nM.}
\label{fig:PKPD-full}
\end{figure}

To fully explore the consequences of accumulation effects on the pharmacodynamics of antimalarial treatment, we simulate the time course of total viable parasite count under a standard AS7 dosing regimen (i.e.\ a dose of 2 mg/kg artesunate every 24 hours for 7 days) for both the 3D7 strain and a hypothetical strain which exhibits a slower rate of stress development during the mid-ring stage. Some key simulation details are provided in the caption of Fig.\ \ref{fig:PKPD-full}. We initiate the simulation with $10^{12}$ parasites per patient with a normally distributed age distribution with mean 10 h p.i. and standard deviation 2 h p.i. \ (Fig.\ \ref{fig:PKPD-full} (inset)). For the laboratory 3D7 strain, the model predicts that effective parasite clearance is achieved immediately following the third dose of artesunate (at 48 hours in the model, Fig.\ \ref{fig:PKPD-full} green curve). In contrast, for the hypothetical strain which exhibits a slower development of the stress response during the mid-ring stage (i.e. $\lambda$ is reduced for this, but no other, stage), we observe a clear and substantial delay in parasite clearance. In detail, the red curve in Fig.\ \ref{fig:PKPD-full} shows the parasitaemia curve for a ``resistant'' strain that has $\lambda = 0.1\ \rm h^{-1}$ for the mid-ring stage (with all other parameters (across all stages) unchanged). This simulation has an \emph{a priori} rationale given previous studies that indicate that field-isolates from Pailin (Western Cambodia) display a reduced sensitivity to artemisinin-based therapies during the ring-stage of infection \cite{Saralambaetal2011,Witkowskietal2013,Dogovskietal2015}. We note that while this simple simulation does not model the process of splenic clearance, its behaviour is consistent with clinical observations of a 1.5 -- 2 times longer time to clearance for resistant strains compared to sensitive strains \cite{Saralambaetal2011}.

\section*{Discussion}

Artemisinin resistance has arisen as the major impediment to the continued success of malaria control programs. With new drugs likely to be some time away from licensure and widespread use, how we maintain the effectiveness of artemisinin-based therapies is an important and urgent problem to resolve. In this paper, we have introduced a novel mathematical model that allows for the detailed investigation of the time-dependent response of \emph{P. falciparum} parasites to the artemisinins. We have established the significant influence of the parasite stress response on killing and incorporated this into a simulation of \emph{in vivo} parasite clearance. This is the first study to our knowledge to incorporate the novel time-dependent drug effect into a fully mechanistic PK--PD model, and constitutes an essential step towards development of a comprehensive framework that can be used to optimise existing dosing regimens.

Validated against detailed \emph{in vitro} experimental data \cite{Klonisetal2013,Dogovskietal2015}, the key feature introduced in our model is the concept of dynamic (accumulating) stress ($S$), and the parameter governing the time scale of that process ($\lambda$). The time-evolution of stress determines the development of the killing rate and therefore the probability of parasite survival (as assessed \emph{in vitro} by viability). This conceptualisation of stress has been shown to not only capture the \emph{in vitro} viability data published in \cite{Klonisetal2013} but also identify the relative contributions of drug concentration and stress response in determining the effective killing effect.

Specifically, the estimate for $\lambda$, which determines the strength of the delay, shows that mid-ring stage 3D7 parasites exhibit a substantially more delayed response to DHA exposure than do other stages. This is consistent with the original analyses presented in Klonis \emph{et al.} \cite{Klonisetal2013}, where a semi-mechanistic (but not dynamic) Cumulative Effective Dose (CED) model was used for interpretation of stage-specific drug effects. Klonis \emph{et al.} also reported an unexpected finding that early-ring parasites exhibited a hypersensitivity to DHA compared to mid-ring parasites. Our model results (Fig.\ \ref{fig:params}) show that a rapid induction of maximal killing (i.e.\ a rapid increase in $S$), rather than a large increase in the magnitude of the maximum killing rate itself, is the primary explanation for this hypersensitivity.       

We also examined through simulation of the full PK--PD model how a cumulative drug response affects \emph{in vivo} parasite dynamics. We considered the standard AS7 artesunate treatment regimen (a dose of artesunate every 24 hours for 7 days). Parasites that are more able to withstand exposure (i.e.\ that have a lower $\lambda$) during the mid-ring stage remain in circulation for a significantly greater period of time, with PD profiles that reflect those from patients infected with resistant strains. These results agree with those we have previously presented using the CED-model (Figure 7 in \cite{Dogovskietal2015}), but we emphasise that the results presented here (Fig.\ \ref{fig:PKPD-full}) arise from a fully mechanistic PK--PD model. This is an important distinction as, by construction, our PK--PD model accounts for the ageing and natural replication dynamics of the parasite population, the time-varying nature of the drug-concentration, and the interaction (killing) between parasite and drug in a self-consistent and biologically realistic way. This provides us with enhanced predictive power when compared to the CED-model, which while empirically useful, was not well-suited to \emph{in vivo} simulation. We emphasise that our \emph{in vivo} simulations provide predictions of the number of viable parasites. However, \emph{in vivo} assays cannot distinguish between viable and non-viable parasites, nor detect sequestered parasites. As such, further advances in experimental assays are required to fully test the predictions from these simulations.

We have referred to the modulatory variable, $S$, throughout the paper as a ``stress''. We have done so to provide guidance as to possible biological interpretations of $S$, but for now an incomplete understanding of the mechanism of action of the artemisinins limits the degree to which our phenomenologically-based model can be correlated with specific biological stresses induced by exposure to artemisinins. Recent work \cite{Wangetal2015,Ismailetal2016b} confirms earlier studies \cite{Meshnicketal1996} suggesting that artemisinins exert their activity by alkylating multiple targets within the parasite. Reports of growth retardation, quiescence and dormancy following artemisinin exposure \cite{Tuckeretal2012,Klonisetal2013,Dogovskietal2015,Hottetal2015} are reminiscent of the cytostatic stress response observed in other organisms \cite{Ronetal2007,Ammetal2014}. Further developments in understanding the mechanistic underpinnings of artemisinin activity are required to further refine our model for ``stress''. The details of any of these processes, were they to be confirmed to be associated with cumulative drug effects, would be able to be incorporated into our model in a straightforward manner through adjustment of the equations governing the time-evolution of $k_{max}$ and $K_c$.

An immediate implication of our model concerns the possible mechanism by which the malaria parasite attains resistance to artemisinin. Drug-resistance is typically characterised by an increase in the drug concentration required (\emph{in vitro} or \emph{in vivo}) to achieve maximal (or half maximal) killing. However, our exploratory analysis suggests that increasing tolerance to stress (i.e.\ reducing $\lambda$) can also underpin drug escape. Indeed, the experimental results from \cite{Dogovskietal2015} in which resistance may be overcome through application of proteasome inhibitors such as Carfilzomib support this possibility. Furthermore, if such a mechanism where at play, then long-lived drugs acting on the same (or a similar) pathway, and subject to the same ``resistance'' mutations would not result in a resistant-phenotype when applied for extended periods (as $S$ would still saturate and a high rate of killing would be achieved). This is precisely the behaviour observed for OZ439 (half-life over 10~hours) in recent experiments \cite{Yangetal2016}.

The model we have introduced, while overcoming restrictions of the standard PK--PD approach and successfully capturing the complex dynamics observed in \cite{Klonisetal2013} for the 3D7 strain, has a number of limitations. Most importantly, Eq.\ \ref{eq:M1a} assumes a drug concentration-independent increase in $S$ while $C>C^*$ and then an instantaneous return to zero when the drug concentration drops below $C^*$. However, this simple approach to modelling the stress, $S$, reflects the current limitations on our understanding of the mechanism of action of the artemisinins. With further data on how the drugs act, the dynamics of stress in the model can be adjusted to reflect the improved understanding. One important avenue to pursue is to examine how parasites that survive exposure to an initial drug pulse respond to a subsequent drug pulse. Does their ``stress'' ($S$) return to zero, or do they display some memory of previous exposure, and so presumably succumb more quickly upon subsequent exposure? If such recovery exists, what is the typical time-scale in relation to the life-cycle? Such possibilities are 1) able to be probed experimentally using the techniques of \cite{Klonisetal2013,Dogovskietal2015}; and 2) able to be readily incorporated into more complex models of the form introduced in this paper. Another area for improvement in the approach taken here is in translating from the \emph{in vitro} to \emph{in vivo} situation. For example, our simple simulations assume there is no killing of parasites due to immune response mechanisms triggered within the host. While the immune response is unlikely to play a major role during the early stages of infection, as infection progresses its effects would be anticipated to become more significant. Therefore, given the fact that both the drug effect and immune response are dynamic in nature, it will be important to explore how differences in the timing of drug application and activation of various immune mechanisms impact upon parasite clearance and optimisation of drug regimens. In the meantime, our results provide new insight into how \emph{P. falciparum} responds to drug. Our model provides an enhanced predictive platform for evaluating the likely efficacy of alternative artemisinin-based drug regimens, directly contributing to the efforts to maintain effective control of malaria.

\section*{Competing interests}

We have no competing interests.

\section*{Author contributions}

PC, NK, LJW, FJIF, LT, JAS and JMM conceived the study. NK, CD, SCX performed the experiment. PC, SS and JMM developed the model. PC and SZ performed the statistical analysis. PC, SZ, JAS and JMM wrote the manuscript. All authors gave final approval for publication.

\section*{Acknowledgments}

We acknowledge many useful conversations with David S. Khoury, Deborah Cromer and Miles P. Davenport (Kirby Institute, UNSW, Australia) and Joel Tarning and Nicholas J. White (Mahidol University, Bangkok, Thailand). Pengxing Cao would like to thank TDModNet for providing a travelling fellowship to enable collaboration on this project.

\section*{Funding information}
The work was supported by the National Health and Medical Research Centre of Australia (NHMRC) through Project Grants 1100394 and 1060357, the Centre for Research Excellence VicBioStat (1035261) and the Centre for Research Excellence PRISM$^2$ (1078068). The Mahidol-Oxford Tropical Medicine Research Unit is supported by the Wellcome Trust of Great Britain. James M. McCaw and Freya Fowkes were supported by Australian Research Council (ARC) Future Fellowships. Leann Tilley was supported by an ARC Professorial Fellowship. Julie A. Simpson was supported by a NHMRC Senior Research Fellowship.

\newpage
\begin{table}[ht!]
\caption*{Table 1: Results of fitting the model to viability data Viability data are for the 3D7 strain \cite{Klonisetal2013}. $\gamma$ is fixed to be 1.7892 based on estimates in Table S1. The model-based 95\% CI and parametric bootstrap 95\% CI are introduced in \emph{Materials and Methods}.}

\begin{center}
\doublespacing
\begin{tabular}{|c|c|c|c|c|}
   \hline
   Parameter (unit) & Estimate & SE & Model-based 95\% CI & Parametric bootstrap 95\% CI \\
   \hline
   \multicolumn {5} {|l|} {\bf Early ring stage}\\
   \hline
   $\lambda$ ($\rm h^{-1}$) & 6.2504 & 0.5745 & (5.1243, 7.3765) &  (1.8587, 9.5201) \\
   \hline
   $\gamma$ & \multicolumn {4} {|c|} {1.7892 (fixed)}\\
   \hline
   $\alpha$ ($\rm h^{-1}$) & 1.6915 & 0.1378 & (1.4215, 1.9616) & (1.2944, 1.8475) \\
  \hline
   $\beta_1$ ($\rm nM$) & 990.84 & 373.49 & (258.81, 1722.9) & (-20377, 1876.5) \\
  \hline
   $\beta_2$ ($\rm nM$) & 12.519 & 1.0631 & (10.435, 14.602) & (10.560, 13.553) \\
   \hline
      \multicolumn {5} {|l|} {\bf Mid-ring stage}\\
   \hline
   $\lambda$ ($\rm h^{-1}$) & 0.3729 & 0.1406 & (0.0974, 0.6485) &  (0.2515, 0.5035) \\
   \hline
   $\gamma$ & \multicolumn {4} {|c|} {1.7892 (fixed)}\\
   \hline
   $\alpha$ ($\rm h^{-1}$) & 1.1224 & 0.2455 & (0.6412, 1.6036) & (0.6934, 1.3371) \\
  \hline
   $\beta_1$ ($\rm nM$) & 224.39 & 112.12 & (4.6466, 444.14) & (117.37, 301.69) \\
  \hline
   $\beta_2$ ($\rm nM$) & $9.97 \times 10^{-4}$ & $1.26 \times 10^{-4}$ & $(7.5, 12.4) \times 10^{-4}$ & $(9.8, 10.1) \times 10^{-4}$ \\
   \hline
      \multicolumn {5} {|l|} {\bf Early trophozoite stage}\\
   \hline
   $\lambda$ ($\rm h^{-1}$) & 1.2290 & 0.2249 & (0.7882, 1.6698) &  (0.6331, 1.8059) \\
   \hline
   $\gamma$ & \multicolumn {4} {|c|} {1.7892 (fixed)}\\
   \hline
   $\alpha$ ($\rm h^{-1}$) & 5.7434 & 0.7460 & (4.2813, 7.2054) & (1.8799, 7.2729) \\
  \hline
   $\beta_1$ ($\rm nM$) & 317.64 & 86.143 & (148.80, 486.48) & (-9.4144, 462.98) \\
  \hline
   $\beta_2$ ($\rm nM$) & 39.570 & 4.6038 & (30.546, 48.593) & (29.315, 46.842) \\
   \hline
      \multicolumn {5} {|l|} {\bf Late trophozoite stage}\\
   \hline
   $\lambda$ ($\rm h^{-1}$) & 2.0906 & 0.2909 & (1.5203, 2.6608) &  (1.4406, 2.6302) \\
   \hline
   $\gamma$ & \multicolumn {4} {|c|} {1.7892 (fixed)}\\
   \hline
   $\alpha$ ($\rm h^{-1}$) & 2.8626 & 0.1591 & (2.5508, 3.1744) & (2.2851, 3.2810) \\
  \hline
   $\beta_1$ ($\rm nM$) & 740.02 & 178.77 & (389.64, 1090.41) & (160.00, 1071.6) \\
  \hline
   $\beta_2$ ($\rm nM$) & 41.405 & 3.6606 & (34.230, 48.580) & (35.459, 45.629) \\
   \hline
\end{tabular}
\end{center}
\end{table}

\newpage
\thispagestyle{empty}

\begin{figure}[ht!]
\centering
\includegraphics[scale=0.8]{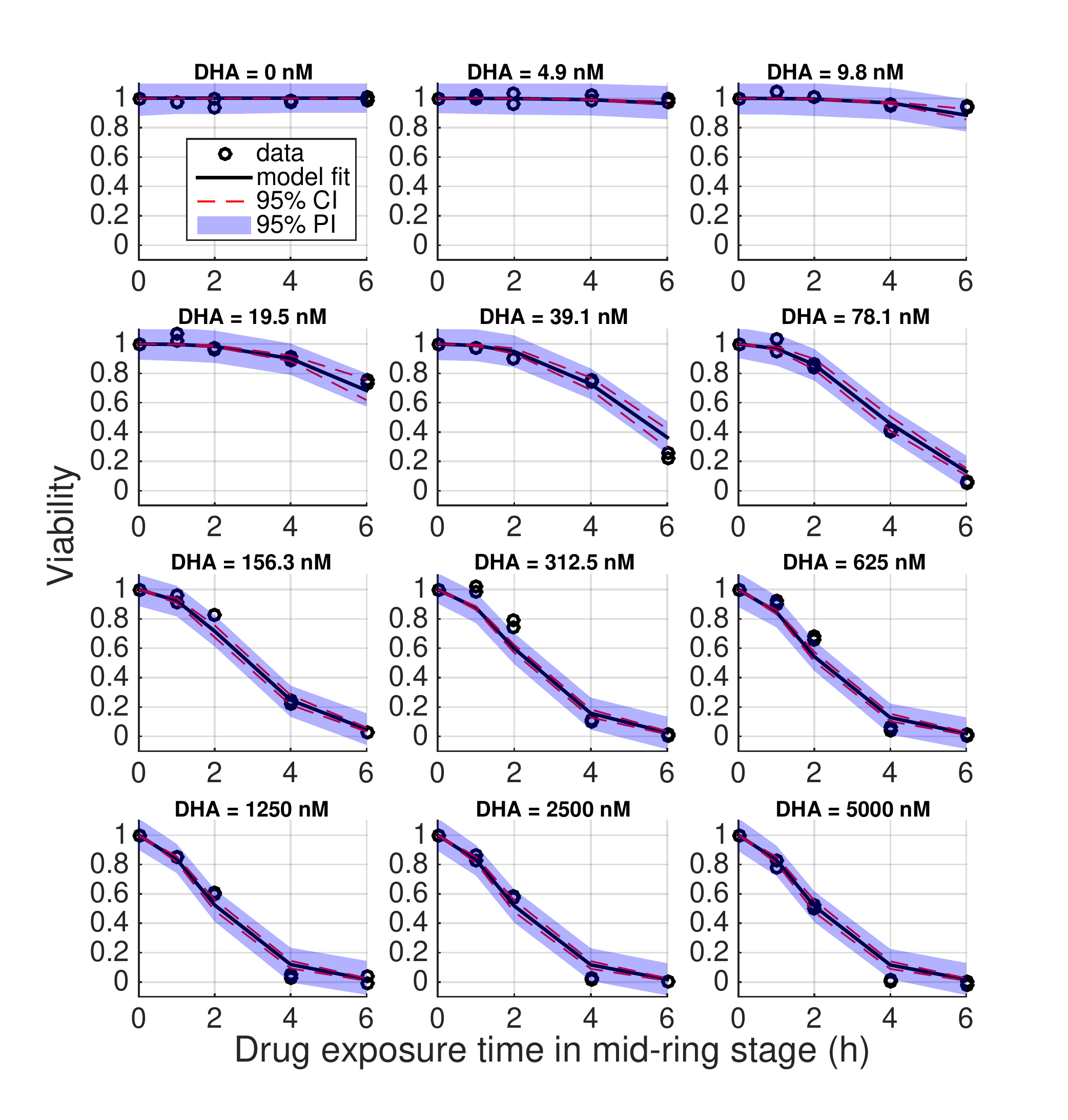}
\caption*{Figure S1: Results of fitting the model to viability data (mid-ring stage). The applied DHA concentration (which then decays) is indicated in the title of each panel. Empty circles are viability data points and duplicate data points for each condition are shown. The black curves are best-fits with 95\% confidence intervals (CI) and 95\% prediction intervals (PI).}
\end{figure}
\newpage
\thispagestyle{empty}

\begin{figure}[ht!]
\centering
\includegraphics[scale=0.8]{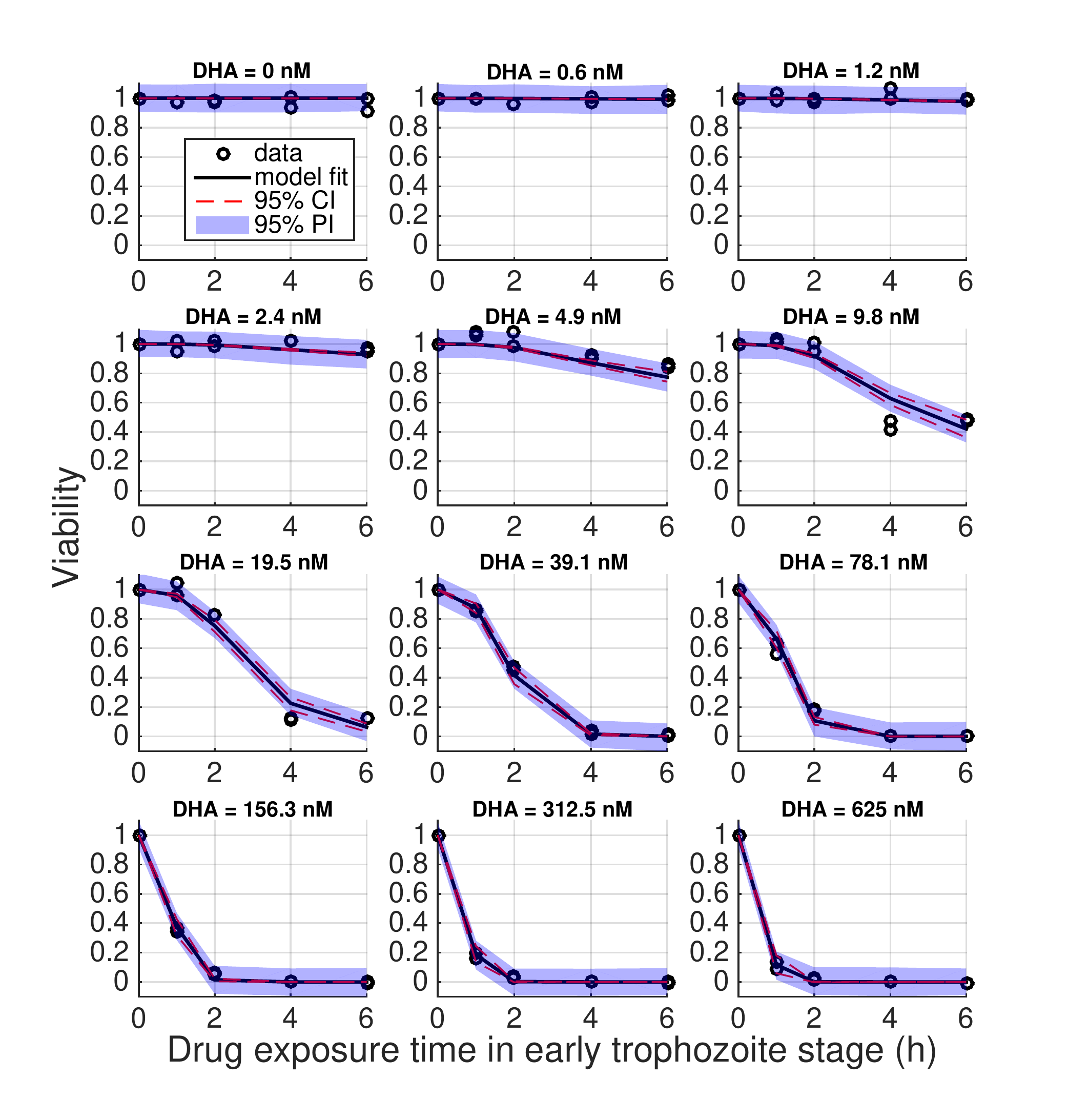}
\caption*{Figure S2: Results of fitting the model to viability data (early trophozoite stage). The applied DHA concentration (which then decays) is indicated in the title of each panel. Empty circles are viability data points and duplicate data points for each condition are shown. The black curves are best-fits with 95\% confidence intervals (CI) and 95\% prediction intervals (PI).}
\end{figure}
\newpage
\thispagestyle{empty}

\begin{figure}[ht!]
\centering
\includegraphics[scale=0.8]{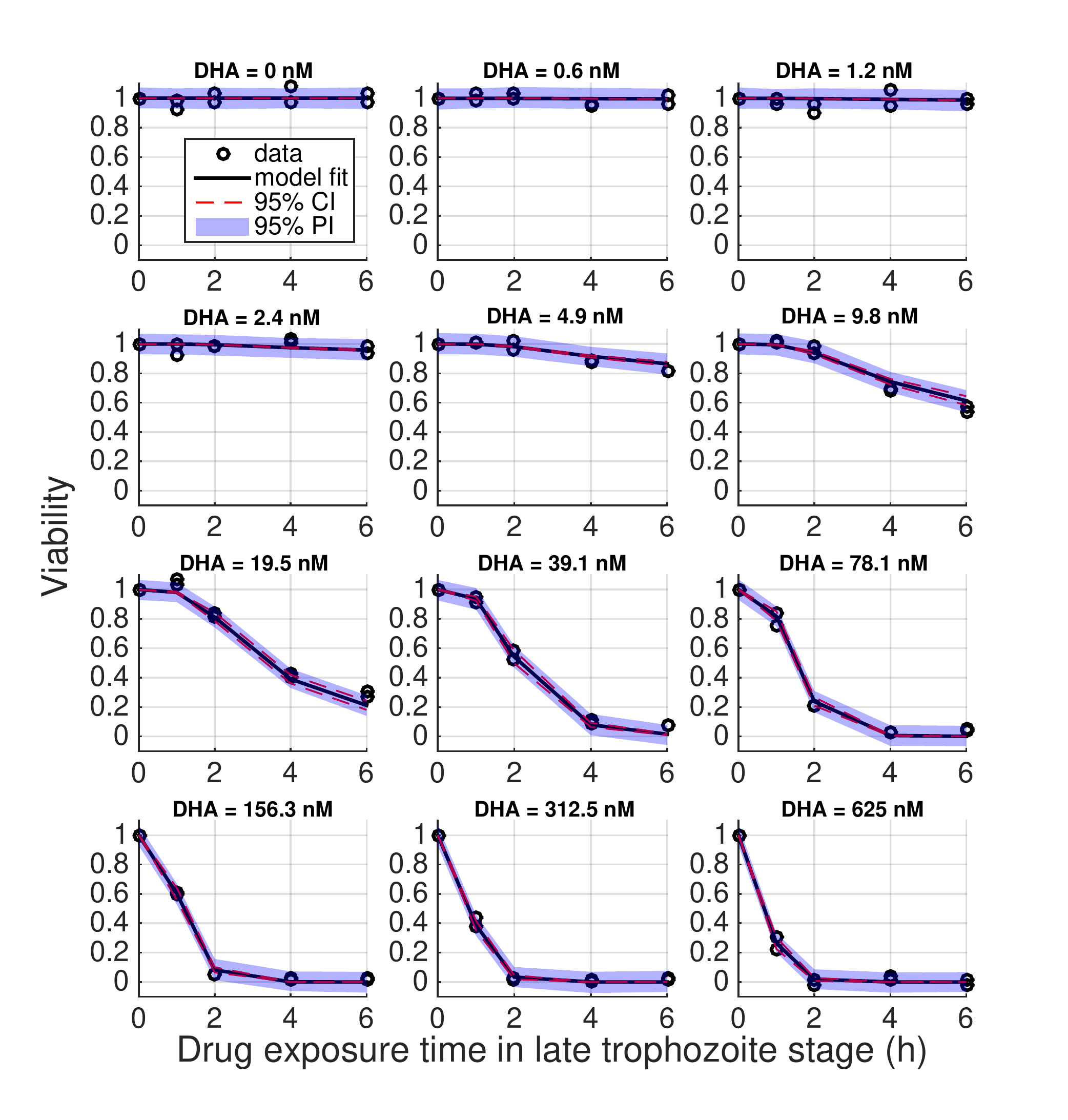}
\caption*{Figure S3: Results of fitting the model to viability data (late trophozoite stage). The applied DHA concentration (which then decays) is indicated in the title of each panel. Empty circles are viability data points and duplicate data points for each condition are shown. The black curves are best-fits with 95\% confidence intervals (CI) and 95\% prediction intervals (PI).}
\end{figure}

\newpage
\begin{table}[ht!]
\caption*{Table S1: Results of fitting the model to viability data. The model-based 95\% CI and parametric bootstrap 95\% CI are introduced in \emph{Materials and Methods} in the main text. $\gamma$ is allowed to vary for different stages.}
\begin{center}
\doublespacing
\begin{tabular}{|c|c|c|c|c|}
   \hline
   Parameter (unit) & Estimate & SE & Model-based 95\% CI & Parametric bootstrap 95\% CI \\
   \hline
   \multicolumn {5} {|l|} {\bf Early ring stage}\\
   \hline
   $\lambda$ ($\rm h^{-1}$) & 6.2913 & 0.5294 & (5.2537, 7.3288) &  (-0.2036, 9.5726) \\
   \hline
   $\gamma$ & 1.7703 & 0.1694 & (1.4383, 2.1022) & (1.5789, 1.9794) \\
   \hline
   $\alpha$ ($\rm h^{-1}$) & 1.6987 & 0.1484 & (1.4078, 1.9896) & (1.0569, 1.8591) \\
  \hline
   $\beta_1$ ($\rm nM$) & 1013.0 & 361.39 & (304.71, 1721.3) & (-76380, 1909.0) \\
  \hline
   $\beta_2$ ($\rm nM$) & 12.734 & 2.0654 & (8.6858, 16.782) & (7.5398, 14.927) \\
   \hline
      \multicolumn {5} {|l|} {\bf Mid-ring stage}\\
   \hline
   $\lambda$ ($\rm h^{-1}$) & 0.3638 & 0.1359 & (0.0965, 0.6301) &  (0.2216, 0.5375) \\
   \hline
   $\gamma$ & 1.7433 & 0.2889 & (1.1770, 2.3096) & (1.1612, 2.0665) \\
   \hline
   $\alpha$ ($\rm h^{-1}$) & 1.1472 & 0.2806 & (0.5973, 1.6971) & (0.3129, 1.4028) \\
  \hline
   $\beta_1$ ($\rm nM$) & 222.94 & 94.830 & (37.076, 408.81) & (113.78, 302.00) \\
  \hline
   $\beta_2$ ($\rm nM$) & $9.86 \times 10^{-4}$ & $4.66 \times 10^{-5}$ & $(8.90, 10.8) \times 10^{-4}$ & $(9.70, 10.0) \times 10^{-4}$ \\
   \hline
      \multicolumn {5} {|l|} {\bf Early trophozoite stage}\\
   \hline
   $\lambda$ ($\rm h^{-1}$) & 1.2720 & 0.2163 & (0.8480, 1.6959) &  (0.8361, 1.7458) \\
   \hline
   $\gamma$ & 2.0864 & 0.2023 & (1.6900, 2.4829) & (1.6895, 2.3859) \\
   \hline
   $\alpha$ ($\rm h^{-1}$) & 4.8326 & 0.6615 & (3.5361, 6.1291) & (2.3239, 5.8950) \\
  \hline
   $\beta_1$ ($\rm nM$) & 280.55 & 72.477 & (138.50, 422.60) & (60.544, 387.98) \\
  \hline
   $\beta_2$ ($\rm nM$) & 26.711 & 6.3515 & (14.262, 39.160) & (13.965, 34.734) \\
   \hline
      \multicolumn {5} {|l|} {\bf Late trophozoite stage}\\
   \hline
   $\lambda$ ($\rm h^{-1}$) & 2.3076 & 0.3986 & (1.5263, 3.0890) &  (1.4744, 2.8986) \\
   \hline
   $\gamma$ & 1.5568 & 0.0799 & (1.4001, 1.7134) & (1.4097, 1.6828) \\
   \hline
   $\alpha$ ($\rm h^{-1}$) & 3.3892 & 0.3381 & (2.7265, 4.0519) & (2.1815, 4.0184) \\
  \hline
   $\beta_1$ ($\rm nM$) & 1132.2 & 469.61 & (211.78, 2052.6) & (-306.63, 1704.5) \\
  \hline
   $\beta_2$ ($\rm nM$) & 60.559 & 11.012 & (38.976, 82.141) & (34.002, 74.003) \\
   \hline
\end{tabular}
\end{center}
\end{table}

\end{document}